\documentclass[conference]{IEEEtran}
\IEEEoverridecommandlockouts
\usepackage{cite}
\usepackage{amsmath,amssymb,amsfonts}
\usepackage{algorithmic}
\usepackage{graphicx}
\usepackage{textcomp}
\usepackage{dblfloatfix}
\usepackage{xcolor}
\usepackage{hyperref}

\def\BibTeX{{\rm B\kern-.05em{\sc i\kern-.025em b}\kern-.08em
    T\kern-.1667em\lower.7ex\hbox{E}\kern-.125emX}}

\usepackage{multirow}
\usepackage{enumitem}
\usepackage{longtable}
\usepackage{soul}
\usepackage{subfigure}

\definecolor{pink}{rgb}{0.9,0,0.9}

\begin{document}

\title{How Do Open Source Software Contributors Perceive and Address Usability?\\ Valued Factors, Practices, and Challenges}

\author{\IEEEauthorblockN{Wenting Wang}
\IEEEauthorblockA{\textit{McGill University}\\
wenting.wang@mail.mcgill.ca}
\and
\IEEEauthorblockN{Jinghui Cheng}
\IEEEauthorblockA{\textit{Polytechnique Mont\'eal}\\
jinghui.cheng@polymtl.ca}
\and
\IEEEauthorblockN{Jin L.C. Guo}
\IEEEauthorblockA{\textit{McGill University}\\
jguo@cs.mcgill.ca}
}

\maketitle

\begin{abstract}
Usability is an increasing concern in open source software (OSS). Given the recent changes in the OSS landscape, it is imperative to examine the OSS contributors' current valued factors, practices, and challenges concerning usability. We accumulated this knowledge through a survey with a wide range of contributors to OSS applications. Through analyzing 84 survey responses, we found that many participants recognized the importance of usability. While most relied on issue tracking systems to collect user feedback, a few participants also adopted typical user-centered design methods. However, most participants demonstrated a system-centric rather than a user-centric view. Understanding the diverse needs and consolidating various feedback of end-users posed unique challenges for the OSS contributors when addressing usability in the most recent development context. Our work provided important insights for OSS practitioners and tool designers in exploring ways for promoting a user-centric mindset and improving usability practice in the current OSS communities.
\end{abstract}


\section{Introduction}
Usability determines how easy, efficient, error-preventing, and pleasant a software system to be used by human users. This quality attribute has been identified as one of the biggest issues that open source software (OSS) has struggled with for more than a decade~\cite{Nichols2003}. 
Distinctive characteristics of OSS development, such as geographically distributed community~\cite{Aberdour2007}, diverse background in members~\cite{Vasilescu2015}, asynchronous communication~\cite{arya2019analysis}, as well as flat organizational structures~\cite{cheng2019activity, Tsay2014}, have shaped the ways how OSS communities consider, identify, and solve usability issues. Understanding how OSS contributors currently address usability in such a distributed collaborative environment will not only help us create an up-to-date overall view of OSS usability practice, but such knowledge will also inform new methods and tools for supporting OSS communities in improving the usability of their projects.

The quest to understanding the issues around OSS usability is not unprecedented. Studies by Andreasen et al.~\cite{Andreasen2006} and Raza and Capretz~\cite{Raza2012} have both identified that while OSS contributors regarded usability as important, it was considered as an ``add-on'' aspect incorporated only during a certain stage of development. In contrast, Terry et al.~\cite{Terry2010} found that OSS contributors possessed a rather ``sophisticated notion of usability'' and have adopted a wide range of practices. Recently, Llerena et al.~\cite{Llerena2019} have also identified obstacles in adapting several usability techniques in OSS projects. Looking back at these previous investigations, it remains unclear as to the overall perception of OSS contributors towards usability, the practicality of the available methods and tools, and the associated challenges.

Meanwhile, the landscape of OSS development has changed drastically over the past decades~\cite{Schrape2017}. Particularly, the user base of OSS applications grew significantly and the composition of the OSS communities has also become increasingly more diverse in terms of gender, tenure, and technical background~\cite{Vasilescu2015, cheng2019activity}. Moreover, tools for OSS development are becoming more abundant~\cite{Trockman2018}. Issue tracking and discussion platforms, where the OSS participants collaboratively consider various aspects of their projects, including usability, are more integrated with the development process~\cite{arya2019analysis,Cheng2018}. Given these drastic changes in the OSS landscape, it is thus difficult for researchers and practitioners to assess the validity of the previous studies and to interpret their implications in the current context of OSS development.

In this article, we report our investigation on the OSS participants' valued factors, practices, and challenges towards usability in the most relevant environment of OSS development. Particularly, we conducted a survey that involved 84 OSS community members who were actively providing diverse contributions to a wide range of projects. With this survey study, we seek to answer the following questions: (1) How do OSS contributors perceive the determining factors associated with usability? (2) What are the prominent approaches OSS contributors currently carry out to address usability issues? (3) What are the prominent barriers and difficulties they faced with respect to addressing usability?

\section{Methods}
\label{sec:methods}

\subsection{Survey Process}
We conducted a survey with OSS contributors who focused on application projects that have a graphic user interface (GUI). The survey was made available from July 1st to August 10th, 2018.

\subsubsection{Instrument}
\begin{figure}[!t]
    \centering
    \includegraphics[width=0.9\linewidth]{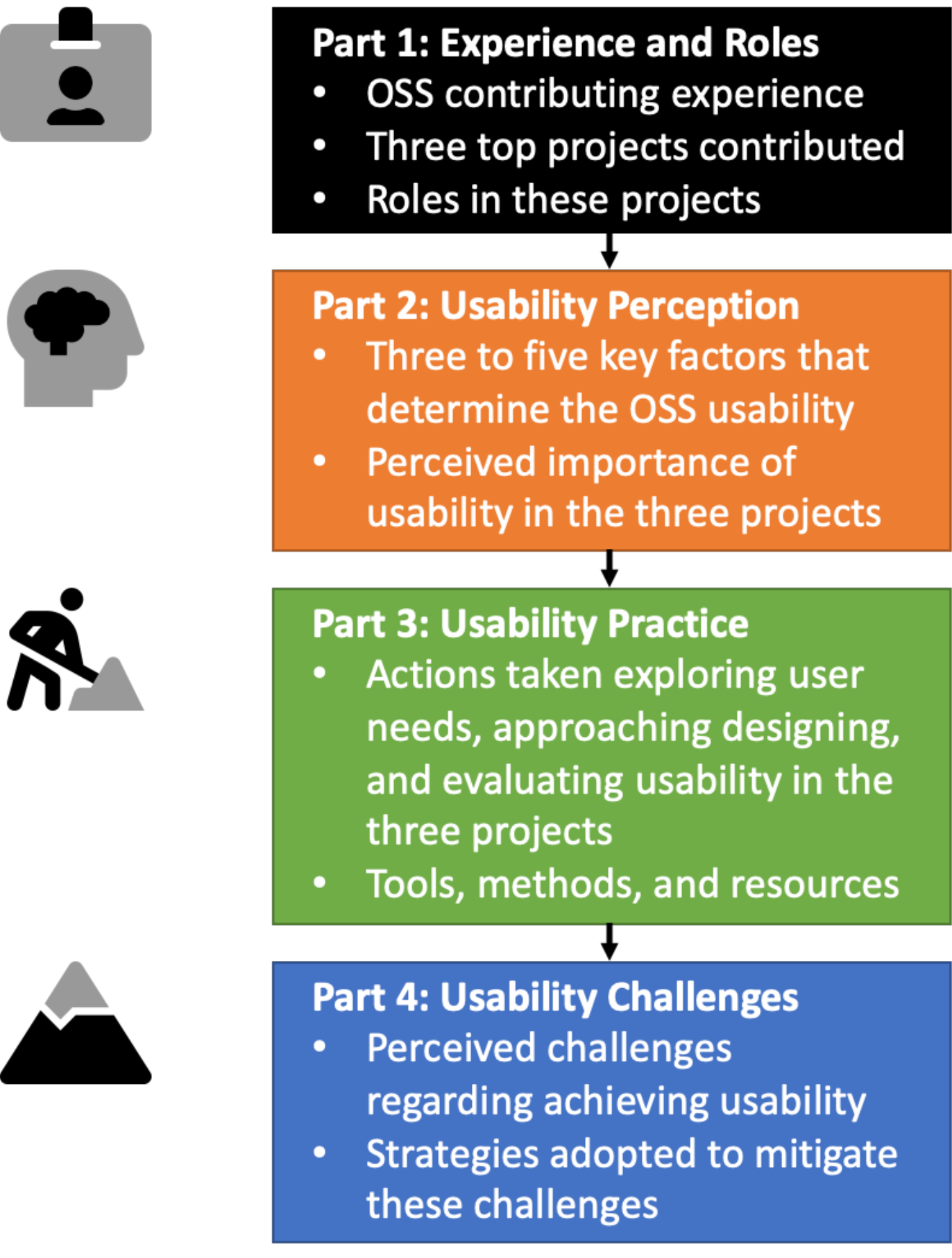}
    \caption{Survey instrument}
    \label{fig:survey}
\end{figure}

We designed the initial survey and improved the survey with a pilot study with Computer Science graduate students. The final survey instrument, which contains mostly open-ended questions, is summarized in Figure~\ref{fig:survey} and available at \href{http://doi.org/10.5281/zenodo.3937877}{http://doi.org/10.5281/zenodo.3937877}. 

\subsubsection{Recruitment}
We first manually compiled a list of 144 popular GitHub projects that had an extensive GUI component from two sources: the GitHub \textit{Collections}\footnote{https://github.com/collections} and a Wikipedia page of OSS systems\footnote{https://en.wikipedia.org/wiki/Listoffreeandopen-sourcesoftwarepackages}. These projects covered diverse application domains, including text editors, email systems, media players, software engineering tools, etc.

For each selected project, we chose the top ten contributors from one of the three aspects: code contribution, reporting and discussing issues, and creating and discussing pull requests. Within the survey active period, we sent three rounds of recruiting emails following the established guidelines about contacting OSS contributors\footnote{https://github.com/ghtorrent/ghtorrent.org/blob/master/faq.md\#contacting-users-for-surveys}. In total, 1180 contributors were contacted.

\subsubsection{Participants}
\label{sec:methods_participants}
We received a total of 162 responses to our survey, among which 84 contained meaningful answers to at least one open-ended question and therefore are included in our data analysis (55 completed all survey questions). 
The participants' experience of contributing to OSS projects ranged from 1 to 27 years ($Median=6$). About one fourth ($N=19$) indicated that they have contributed to more than 30 OSS projects; the rest contributed in a median of five projects. These projects are summarized in \href{http://doi.org/10.5281/zenodo.3937877}{http://doi.org/10.5281/zenodo.3937877}).

While around half of the participants reported regular occupations as programmers or software engineers, the rest varied and included quality engineers, managers, designers, students, and domain specialists such as architects, musicians, writers, data scientists, and security experts. Most participants provided multiple types of contributions to OSS projects. Collectively, their contributions are wide-ranging, covering feature development, code maintenance, bug fixing, issues reporting and discussion, user interface design, documentation, quality assurance, user support, public relation, release management, and localization.

\subsection{Response Analysis}
To analyze the survey responses, we followed an inductive thematic analysis process~\cite{saldana_coding_2009}. 
After individual open coding, all authors met to discuss and merged their results while resolving ambiguity and disagreements. 
The first author then re-coded the survey responses based on the agreed coding schema.

\subsection{Threats to Validity}
Despite our effort in recruiting active OSS contributors to a wide range of projects, our recruitment only achieved a response rate of 13.7\%, with only 7.1\% of contacted contributors answered the open-ended questions (comparable with similar studies~\cite{Gousios2016}). Additionally, we only examined the perspectives of contributors to GitHub projects. Future work focusing on other OSS platforms as well as passive users and information consumers within OSS communities can extend our findings. However, we argue that our sample allowed us to expose prominent issues about OSS usability directly from active contributors.
\section{Results: Valued Factors}
We identified six themes from the practitioners' perspective about the most important factors that determine OSS usability.

\subsubsection{Usability-Related System Characteristics}
Most participants ($N=63$) valued certain system-centric quality attributes associated with usability, such as efficiency, consistency, and customizability. When discussing these aspects, the participants usually did not mention the characteristics of the end-users. For example, P20 valued ``\textit{good UI organization (placement of buttons, of settings, etc...).}'' Participants seemed to have paid special attention to usability guidelines for achieving these system-centric qualities. 

\subsubsection{Other System Characteristics}
About half of the participants ($N=41$) also mentioned system characteristics that are only implicitly related to usability (e.g. reliability, maintainability, and feature completeness) as important considerations. Interestingly, some participants mentioned that code quality can also affect usability. As P82 explained, ``\textit{good code will make it easy to improve the software and make it become good too.}''

\subsubsection{User Support}
Some participants ($N=30$) considered quality user support from the development team as an important factor. Such support was usually done on issue tracking systems. For example, P54 mentioned that the usability of their project is constantly improving because the development team ``\textit{responds to users' issues quickly to develop solutions to satisfy the needs of the project.}'' Some participants also considered having a clear policy about user support and developer contribution as an important factor to improve OSS usability.

\subsubsection{Environment and Market}
Participants ($N=26$) also mentioned environmental factors, such as the audience size, the performance of the competitors, and the impact of the system to the users and the society, that have influenced the efforts participants put into improving OSS usability. For example, P65 made an interesting statement about a virtuous cycle: ``\textit{This is a chicken-and-egg problem: more popularity implies more users, ... which implies more questions and answers and better UX, ... which attracts more users, thus increasing popularity. }''

\subsubsection{User Characteristics}
Only about one-thirds of practitioners ($N=25$) indicated that a good understanding of their target users' characteristics (e.g. technical background, domain knowledge, product familiarity) could allow them to design for better usability. For example, P68 focused on retaining power users (i.e. users with advanced knowledge and experience of the system) because they ``\textit{can be highly sensitive to even a small disruption to their workflow or a non-optimal design choice.}'' A few participants also considered users with disabilities or from a different culture to accommodate diverse users.

\subsubsection{Resource and Mindset}
Some participants ($N=17$) believed that resources, such as financial support, time, and expertise, were an important factor for OSS usability. For example, P58 emphasized that it is important to ``\textit{have at least one dedicated designer on the team.}'' Some survey participants also explicitly considered the focus of usability as a top-down pursuit that relies on the opinion and motivation of the core contributors of a project.

\section{Results: Practices}
\label{sec:resPractices}
We examined practitioners' practices in the following activities: (1) exploring users' needs, (2) approaching user interaction design, and (3) evaluating the usability. Although the boundaries are sometimes blurred, we separated them in our survey questions and data analysis to elicit more detailed information.

\subsubsection{Explore Needs}
Most participants ($N=64$) discussed approaches that welcomed users' opinions when exploring user needs. Among such approaches, however, the initiator of the user involvement differed and included: (1) most commonly, the general users, usually through reporting usability issues in the issue tracking systems; (2) the power users who directly involved in the development process (e.g. through participatory approaches); and (3) the developers themselves, by actively discussing with users about their needs.

On the other hand, 24 participants also reported ways of exploring user needs in which no end-users were directly involved. Among them, only three participants reported cases where they got inspirations from research literature, competitor systems, or opinions of design specialists. Most participants in this group relied on their own experiences or understandings of the target users (e.g. ``\textit{think in the shoes of the user}'' (P5)), sometimes with guilt or regret.

\subsubsection{Approach Design}
Some participants ($N=22$) reported approaches in which users' opinions and feedback were continuously considered during the design process. A trial and error strategy is often used; i.e. user feedback was gathered, again, usually through issue tracking systems, after a design had been implemented in the system and delivered to the users. Only a few participants discussed an alternative approach; e.g., P25 mentioned the benefit of a participatory approach: ``\textit{We directly contact the user during development. ... Once the users get to know that they have a saying in the final result, they get incredibly proficient at giving the right feedback.}''

A total of 30 participants reported cases where they did not include users in the process of design. Many of them ($N=16$) approached the design only based on their own opinions. The others got inspiration from sources such as guidelines and tutorials ($N=7$), usability experts ($N=6$), and competitor products ($N=3$). For example, P11 mentioned that they ``\textit{had a design lead who reviewed all functionality/PRs [pull requests] that had a UI component.}''

\subsubsection{Evaluate Usability}
Although 43 participants mentioned that they evaluated usability with the end-users, most only vaguely discussed their efforts of ``\textit{gathering user feedback},'' often, again, through issue tracking systems. Some participants have discussed informal user evaluation relying on personal connections, telemetry data, or based on direct feedback from power users. Notably, however, there were a few participants ($N=4$) had adopted traditional usability methods, usually supported by people who have usability expertise. For example, P58 described a case in which UX students were involved in the usability evaluation process, an approach that has resulted in ``\textit{valuable results.}''

On the other hand, 17 participants discussed methods that did not involve users. These approaches were mostly based on heuristics and internal evaluation of the development team. The rationale was usually that the developers were also the users. But it can sometimes lead to oversimplified statements. For example, P26 mentioned: ``\textit{If it works for the developers, it works for the user.}''

\section{Results: Challenges}
Participants expressed four major challenges toward resolving OSS usability. For each challenge, they also collectively offered strategies to overcome it.

\subsubsection{Mindset}
Participants ($N=19$) have voiced concerns that many developers are interested in system and feature-related issues, and overlook usability issues. For example, P74 stated that, in their community, ``\textit{UX is perceived as less challenging and interesting compared to exploring new programming technologies.}'' Due to this system-centered mindset and limited resources, project owners often allocate resources on other competing goals (e.g. the desired feature set or system performance) before improving the usability. Interestingly, three participants also mentioned that OSS users' tolerance for bad design had led to lowered motivation for the development team to solve usability issues.

To address this challenge, participants emphasized top-down influences on the value of user feedback and usability considerations. As an example, P58 discussed their strategies of building a design-centered culture: ``\textit{From the very beginning building a strong design culture. Having a "design" tag in the issue tracker. Requiring mockups for new features first before development. Building a design team.}''

\subsubsection{Understanding User Needs}
Participants ($N=17$) discussed that the actual user needs of OSS applications are usually nebulous and evolve frequently. For example, P22 stated: ``\textit{Due to the nature of open source, ... the user data received is from a very small portion who are using the final compiled binaries.}'' Many participants also find it difficult to understand the meanings and rationales of user feedback and requests in issue tracking systems.

To conquer these challenges, participants emphasized the effort of digging deeper into the user feedback. For example, P20 mentioned: ``\textit{[we] pay attention not just to the bug reported but why it was reported and how to best meet the needs of the user.}'' Some practitioners also decided to include more direct user-centered communications and user studies. For example, P22 addressed this challenge ``\textit{by being always available to interact with issues/chat/comments, and by constantly conducting small surveys to collect user opinions.}'' Some participants wished that there was more ``\textit{instrumentation to gather empirical data}'' (P11) in the process to assist understanding of users' needs.

\subsubsection{Addressing User Diversity}
Participants ($N=15$) also mentioned challenges in addressing the needs and opinions of an increasingly diverse user base. After all, features need to be carefully prioritized to avoid, as P75 stated, ``\textit{feature creep from trying to please everyone}.'' UI/UX design in consideration of diverse user characteristics is also a challenge, as P9 shared: ``\textit{When introducing something, we need to be careful in balancing effectiveness for some vs. simpleness for others.}''

To overcome these challenges, participants emphasized the importance of engaging the user community to have a thorough discussion about design before implementation. In terms of design solutions, participants valued customizable user interfaces. However, such features usually add system complexity and rely on the familiarity of the workflow, and thus often intimidate novice users.

\subsubsection{Development Issues}
Participants ($N=13$) also shared the difficulties they encountered during the process of development that have affected usability. These challenges were associated with framework and library limitations, project complexity, and the lack of transparency in the development process. For example, P64 described a long-standing project: ``\textit{Lots of generations of software using different toolkits and libraries... This prevent having a cohesive and well-defined interaction that users can become familiar with.}''

To overcome these development-level challenges, participants advocated practices of taking time to understand the system, the project, and the community culture, before contributing to the project. For example, P78 suggests potential contributors to ``\textit{take a long time to read docs and code}'' before getting involved.
\section{Discussion}
\label{sec:discussion}

\begin{figure*}[!t]
    \centering
    \includegraphics[width=0.9\linewidth]{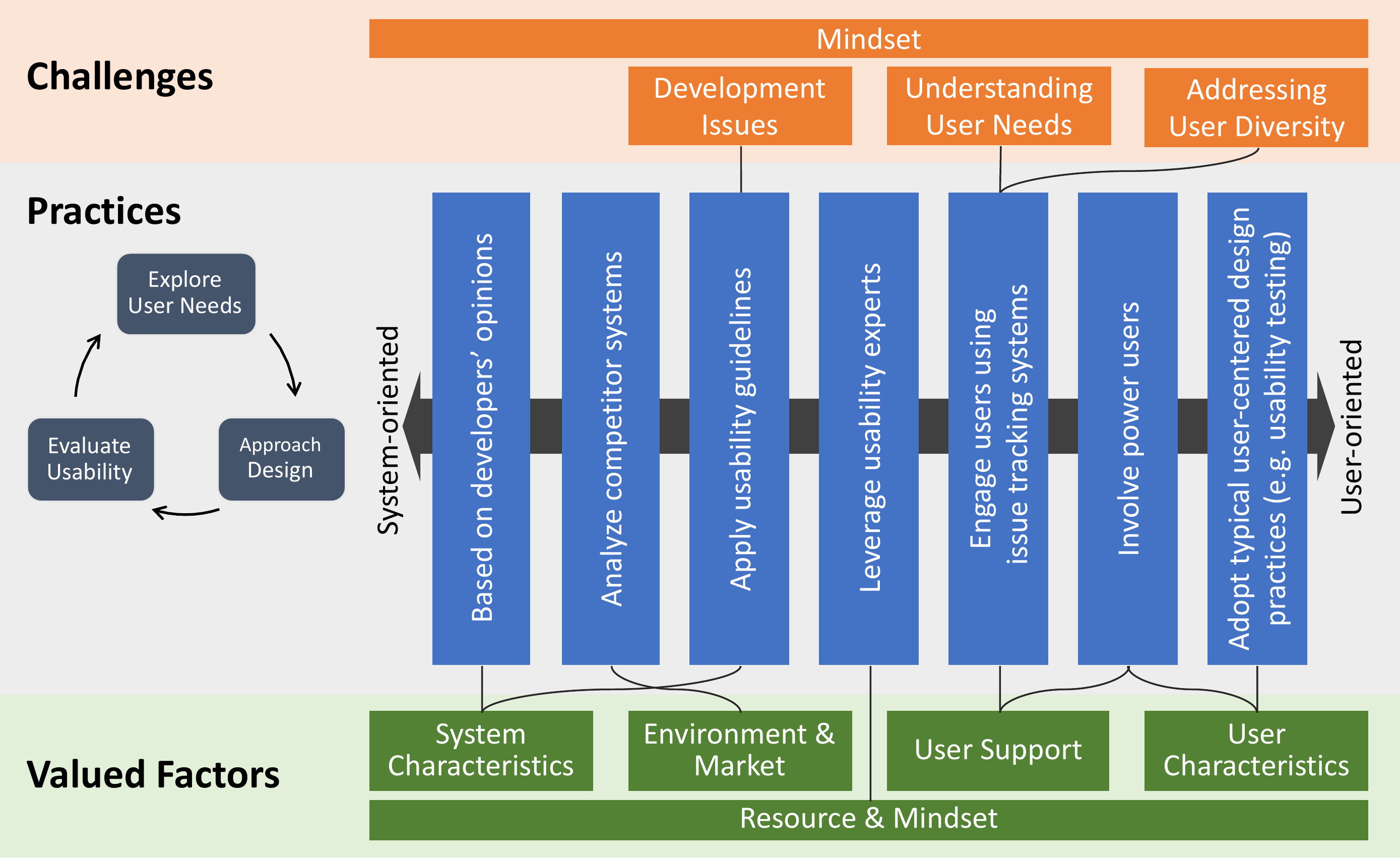}
    \caption{Summary of the valued factors, practices, and challenges of OSS contributors towards usability. Participants' valued factors affected their usability practices. Certain popular practices have resulted in prominent challenges.}
    \label{fig:summary}
\end{figure*}

We aimed to understand the valued factors, practices, and challenges with respect to OSS usability through the active contributors' perspective; Figure~\ref{fig:summary} summarizes our findings. Although some of our participants have adopted typical usability methods, OSS contributors in general tended to associate software usability to system characteristics rather than characteristics, needs, or preferences of end users. Additionally, even with the modern OSS tools that shortened the distance between users and developers, the OSS communities are still facing significant challenges in understanding and addressing user needs. Here we discuss our findings and suggest practical and research directions, derived from our results, that can be helpful to improve the OSS usability practice.

\subsection{Supporting a Deeper Engagement with a Diverse User Base}
Our participants valued providing quality user support and discussed accessibility and globalization issues. Participants also involved users frequently in their practice, majorly through issue tracking systems. Our findings indicated that a balanced relationship between OSS developers and users, in which the former made efforts to actively seek users' opinions, has led to better usability practices.

During their attempts to engage with users, however, participants faced new challenges because of the increasing heterogeneity of the user base. First, users' feedback is usually very diverse and sometimes even contradictory to each other. So the OSS contributors experienced difficulties with respect to (\textit{Addressing User Diversity}) when prioritizing features and choosing design alternatives. Second, the users who provided feedback only represent a small sample of the target users. This is particularly problematic if the development teams solely rely on the passive report mechanism of the issue tracking systems. Indeed, \textit{Understanding User Needs} is a challenge mentioned by many participants.

These findings indicated that supporting OSS practitioners to communicate with and therefore understand their users in the new OSS development context is an important research direction. Particularly, enhancing the current issue tracking systems to help the OSS developers effectively understand the target users and their needs is especially useful. For example, semi-automated tools that summarize the various user comments and generate structured use cases can help developers understand the overall trend of the users' needs to prioritize their work. Tools that support active approaching the end-users and collecting their characteristics can also be helpful.

\subsection{Promoting Motivation for Usability at the Community Level}
The OSS communities now include contributors with a diverse background and different expertise; this is also reflected in our participants' composition. Echoing this new context, participants valued the importance of community and environmental issues when considering usability. Although rare, they also discussed cases where usability experts were involved in various practices. However, many OSS developers still have the \textit{Mindset} of prioritizing system-related issues over usability issues, which poses a major challenge. As such, fostering a welcoming and motivating environment for diverse OSS contributors to understand and discuss usability issues is an important practice and research direction.

A repeated theme in our results is that user adoption (and thus user feedback) and usability of the system grow hand-in-hand. As identified in Terry et al.~\cite{Terry2010}, user adoption and feedback serve as a social rewards factor that promoted contributors' usability practices. In fact, many company or organization-supported OSS projects thrived because of this virtuous cycle. However, we also found that some OSS projects, especially small and volunteer-driven projects, struggled at simply breaking into this growth cycle. Such challenges may only be addressed by a multifaceted solution that involves adjustments in the mindsets of OSS communities as well as the creation of new tools for supporting OSS usability.

Our results suggested several directions in which OSS tools can be improved or developed to contribute to this process. First, as suggested by our participants, simply adding some usability-related features (e.g. adding ``usability'' as one of the default tags in issue tracking systems) may bring community awareness. Second, criteria in the OSS tools for assessing individual contributions can be modified to give users credits for providing high-quality feedback and encourage developers to consider interaction design issues. Third, user-project matching mechanisms can be developed to help the new and/or struggling OSS projects enter the virtuous cycle of user adoption and growth in usability.

\subsection{Enhancing Tools and Techniques for OSS-Oriented User-Centered Design}
Abundant automated tools have been developed to facilitate the continuous delivery of new software features and solutions. However, tool support for streamlining the delivery of user values is lacking. This gap is demonstrated when our participants voiced the challenges in understanding user needs and usability-related development issues. At the same time, our findings revealed that users are usually involved before and after the development process, but not during.

While we have found some participants adopted user-centered design techniques such as low-fidelity prototyping, these cases were still not common and were only associated with bigger projects. Unawareness or difficulties in adopting these techniques in smaller OSS communities were a repeated theme in our survey responses. Lack of knowledge and expertise challenged our participants. The involvement of usability experts in these communities is still rare.

These results indicated several areas in which OSS tools can encourage developers to more effectively adopt user-centered design methods and techniques. First, tools and techniques integrated with the current OSS tool-chain to support the continuous involvement of users throughout the development process will be helpful. For example, new tools that support solicitation of continuous user feedback on low-fidelity prototypes during the design phases can be integrated into the issue tracking systems. Second, our participants have indicated that, although seemingly unrelated, code quality can affect the overall quality of the system, including usability. Thus static analytic tools that detect potential usability-related issues in the code (e.g. inconsistent use of UI libraries) could help resolve certain challenges. It is important to note that tools supporting usability practices should, on one hand, accommodate the communities' common focus on system characteristics, and on the other hand, remove the barrier of contribution from UI/UX experts and end-users.
\section{Conclusion}
This survey study accumulated information about the valued factors, practices, and challenges of OSS contributors towards usability under the most recent development context. Although some of our participants adopted typical usability methods, the general mindset of the OSS contributors was still system-centric rather than user-centric. Additionally, they faced new challenges in understanding their users and consolidating the diverse user feedback. Our results provided valuable insights for OSS practitioners and tool designers in exploring ways to support usability practice in OSS communities.

\section{Acknowledgement}
We thank our participants for their thoughtful answers. This work was funded by the Canada NSERC Discovery Grant [RGPIN-2018-04470].

\bibliographystyle{IEEEtran}
\bibliography{OSSUsability,bibs}

\end{document}